\documentclass[pra,aps,twocolumn,10pt,showpacs,groupedaddress,superscriptaddress,floatfix,notitlepage]{revtex4-2}
\usepackage{amsmath,amsfonts,amssymb,graphics,graphicx,epsfig,color,times, braket}%,bbm}
\usepackage[utf8x]{inputenc}
\usepackage{color}
\usepackage{bbm, dsfont}
\usepackage{subfigure}
\usepackage{hyperref} 
\usepackage{mathrsfs}
\usepackage{verbatim} 
\begin{document}

\title{Manipulating Excitation Dynamics in Structured Waveguide Quantum Electrodynamics}

\author{I Gusti Ngurah Yudi Handayana}
\thanks{These two authors contributed equally}
%\email{ngurahyudi@unram.ac.id}
\affiliation{Molecular Science and Technology Program, Taiwan International Graduate Program, Academia Sinica, Taiwan}
\affiliation{Department of Physics, National Central University, Taoyuan City 320317, Taiwan}
\affiliation{Institute of Atomic and Molecular Sciences, Academia Sinica, Taipei 10617, Taiwan}

\author{Ya-Tang Yu}
\thanks{These two authors contributed equally}
%\email{yatan1018@gmail.com}
\affiliation{Institute of Atomic and Molecular Sciences, Academia Sinica, Taipei 10617, Taiwan}

\author{Wei-Hsuan Chung}
%\email{b12202069@ntu.edu.tw}
\affiliation{Department of Physics, National Taiwan University, Taipei 10617, Taiwan}
\affiliation{Institute of Atomic and Molecular Sciences, Academia Sinica, Taipei 10617, Taiwan}

\author{H. H. Jen}
\email{sappyjen@gmail.com}
\affiliation{Institute of Atomic and Molecular Sciences, Academia Sinica, Taipei 10617, Taiwan}
\affiliation{Molecular Science and Technology Program, Taiwan International Graduate Program, Academia Sinica, Taiwan}
\affiliation{Physics Division, National Center for Theoretical Sciences, Taipei 10617, Taiwan}

\date{\today}
\renewcommand{\r}{\mathbf{r}}
\newcommand{\f}{\mathbf{f}}
\renewcommand{\k}{\mathbf{k}}
\def\p{\mathbf{p}}
\def\q{\mathbf{q}}
\def\bea{\begin{eqnarray}}
\def\eea{\end{eqnarray}}
\def\ba{\begin{array}}
\def\ea{\end{array}}
\def\bdm{\begin{displaymath}}
\def\edm{\end{displaymath}}
\def\red{\color{red}}
\pacs{}

\begin{abstract}
Waveguide quantum electrodynamics (wQED) has become a central platform for studying collective light–matter interactions in low-dimensional photonic environments. While conventional wQED systems rely on uniform chirality or reciprocal emitter–waveguide coupling, we propose a \textit{structured wQED} framework, where the coupling directionality of each emitter can be engineered locally to control excitation transport in an atom--nanophotonic interface. For different combinations of patterned coupling directionalities of the emitters, we identify four representative configurations that exhibit distinct dynamical behaviors---centering, wave-like, leap-frog, and dispersion excitations. Spectral analysis of the effective non-Hermitian Hamiltonian reveals that these dynamics originate from interferences among subradiant eigenmodes. Variance analysis further quantifies the spreading of excitation as functions of interatomic spacing and global chirality, showing tunable localization–delocalization transitions. Including nonguided losses, we find that the transport characteristics remain robust for realistic coupling efficiencies ($\beta \geq 0.99$). These results establish structured wQED as a practical route to manipulate excitation localization, coherence, and transport through programmable directionality patterns, paving the way for controllable subradiant transport and chiral quantum information routing. 
\end{abstract}
\maketitle

%%%%%%%%%%%%%%
\section{Introduction}
Waveguide quantum electrodynamics (wQED) has emerged as a central platform for exploring collective radiations and photon-photon correlations \cite{Sheremet2023}, supported by advances in nanofabrication techniques and photonic technologies that enable precise control over emission and coupling processes \cite{Vetsch2010, Thompson2013, Goban2015, Corzo2019, Kim2019, Dordevic2021}. In these atom-nanophotonic systems \cite{Mitsch2014, Chang2018, Kim2021}, quantum emitters couple collectively to guided photonic modes \cite{Douglas2015, Solano2017,Tudela2024}, enabling long-range and collective interactions \cite{Jen2025}, giving rise to phenomena such as collective radiative dynamics \cite{Henriet2019, Zhang2019, Ke2019, Albrecht2019, Needham2019, Mahmoodian2020, Masson2020,Jen2021_bound, Pennetta2022, Pennetta2022_2}, photon-mediated \cite{Zhong2020} or disorder-induced localization \cite{Mirza2017, Fayard2021, Wu2024} and photon correlations \cite{Tian2025}, excitation trapping effect \cite{Handayana2024}, distinctive quantum correlations \cite{Tudela2013, Mahmoodian2018, Jeannic2021,Jen2022_correlation,Handayana2025}, and graph states generation \cite{Chien2023, Goswami2025}. These features make wQED a versatile setting for studying nonequilibrium phase transitions \cite{Diehl2010} and for controlling light-matter correlations in open and low-dimensional photonic environments \cite{Tudela2024, Kien2005}.

\begin{figure}[t]
	\centering
	\includegraphics[width=0.49\textwidth]{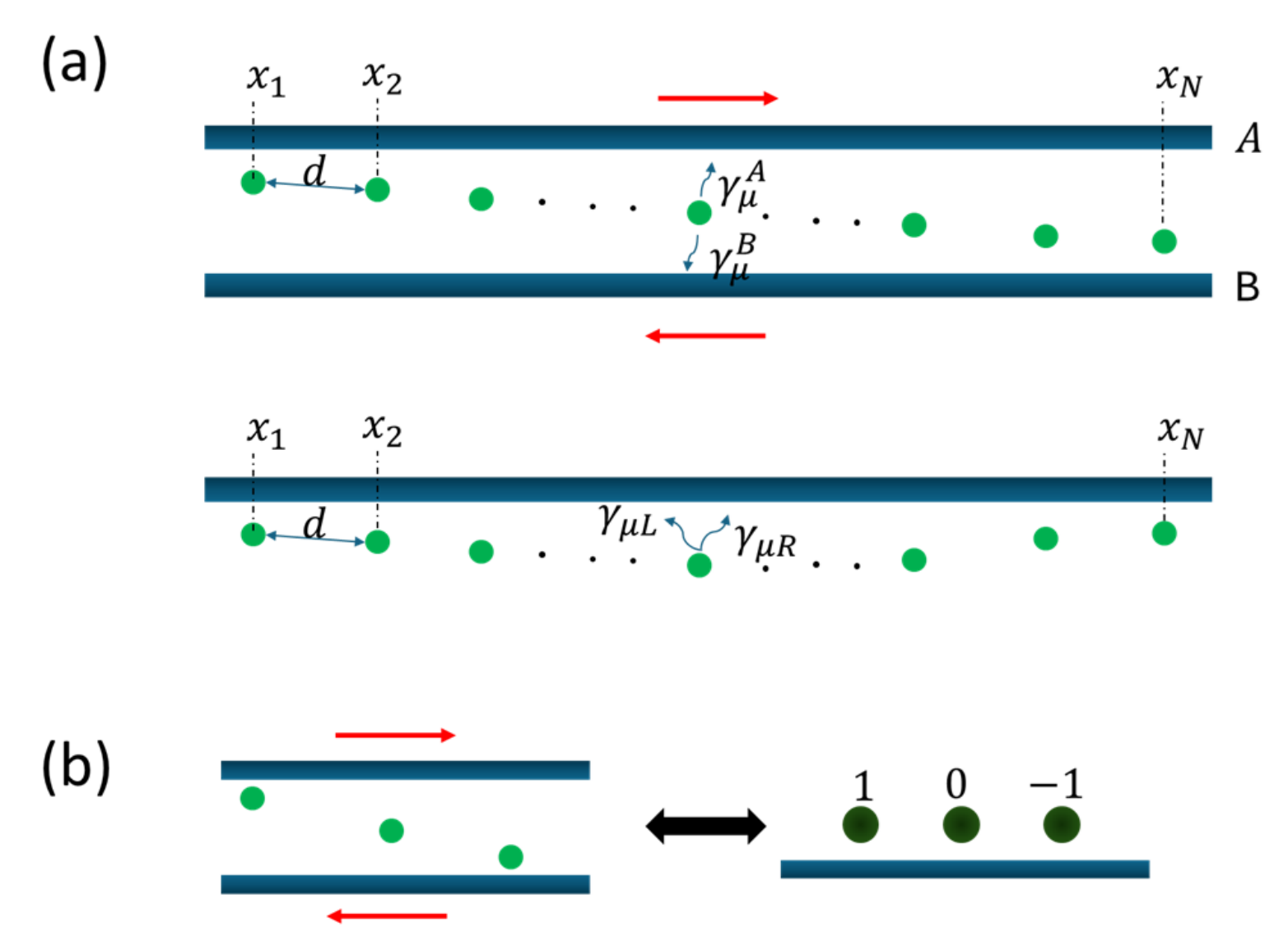}
	\caption{(a) Equivalence between a parallel double-waveguide system (upper panel) and an engineered single-waveguide configuration (lower panel). In the double-waveguide configuration, waveguides A and B support opposite unidirectional propagations (red arrows), while the atoms are positioned at an angle with respect to each other, leading to nonuniform coupling strengths and unequal decay rates $\gamma_{\mu}^A$ and $\gamma_{\mu}^B$ into the respective waveguides. In the equivalent single-waveguide mapping, the same effective dynamics can be reproduced by tailoring the transverse positions of atoms to control the coupling strength and by setting the local directionality. (b) An example to show the equivalence between double-waveguide and structured wQED systems where the two configurations (left and right panels) share an identical directionality pattern. In this example, the directionality is varied site-by-site with unidirectional ($D_\mu = 1$) and reciprocal ($D_\mu = 0$) emitters indicated explicitly for clarity.}\label{Fig1}
\end{figure}

One of the most distinctive features emerged in the wQED setting is chirality \cite{Pichler2015,Lodahl2017}—where the time-reversal symmetry is broken, so that photons preferentially travel in one direction \cite{Gardiner1993,Carmichael1993,Stannigel2012}.
Chiral light–matter coupling enables nonreciprocal phenomena such as cascaded quantum emissions \cite{Jen2020_collective} and direction-dependent excitation transport \cite{Holzinger2024,Diepen2025,Chen2025, Yu2025}. Recent reviews on chiral quantum optics have highlighted its implementation across various photonic platforms—from 0D cavities to 2D nanostructures—and the growing importance of controlling light–matter coupling directionality \cite{Forero2025}. Efforts to realize this asymmetry have been demonstrated in double-waveguide \cite{Yang2022} or single-waveguide configurations \cite{Lee2025}. The former uses parallel waveguides to support unidirectional propagation with atoms displaced in between the waveguides, while the latter uses a single waveguide but the atomic positions follow a specific geometric pattern to tailor the coupling strength [Fig. \ref{Fig1}(a)]. Both schemes, however, either rely on uncontrolled propagation mode ratios among emitters or require a uniform directionality imposed across the entire array, lacking a controllable and flexible knob on the coupling directionality. In contrast, the spin–momentum locking enables each emitter to be controlled individually through its Zeeman sublevels \cite{Mitsch2014,Lodahl2017}, allowing each atom to exhibit distinct left–right asymmetry—effectively realizing behaviors analogous to double-chiral-waveguide and structured waveguide configurations within a single controllable framework [Fig. \ref{Fig1}(b)].

This structured wQED framework, building upon the concept of locally controllable emission, allows the directionality of each emitter deliberately arranged in space to form a prescribed pattern. Rather than assuming a uniform or globally chiral configuration \cite{Stannigel2012,Jen2020_collective,Yu2025}, each atom possesses its own coupling ratio to the right- and left-propagating modes \cite{Pichler2015}, collectively defining a directionality pattern. Table \ref{table1} provides a concise comparison between conventional schemes and the structured wQED framework introduced here. Although conceptually straightforward, this structured wQED system remains largely unexplored. Here, we address this gap by formulating a generalized framework and identifying four representative directionality patterns that give rise to distinct excitation behaviors—centering, wave-like, leap-frog, and dispersion excitation—emerging beyond the concentrated excitation dynamics \cite{Yang2022}. We further characterize these behaviors through spectral analysis of the system’s non-Hermitian eigenmodes, revealing clear connections between long-lived subradiant states and the observed transport features.

\begin{table*}[t]
	\centering
	\caption{Comparison between typical waveguide QED schemes and the structured wQED framework considered in this work.}\label{table1}
\begin{tabular}{lccc}
		\hline\hline
		& Uniform chirality & Double-waveguide & Structured wQED \\
		\hline
		Scale of control & Global & Global or geometric & Site-dependent $D_\mu$ \\
		Limitation of chiral transport & Fixed by system & Limited & Fully programmable \\
		Spatial control of transport & Not accessible & Partial & Designed by pattern \\
		\hline\hline
	\end{tabular}
\end{table*}

These findings provide direct evidence that structured wQED introduces an additional degree of freedom, enabling precise control of excitation dynamics in waveguide-coupled quantum systems. We further evaluate its robustness against coupling imperfections, highlighting the experimental feasibility of realizing these dynamics in state-of-the-art nanophotonic platforms such as superconducting qubits \cite{Roushan2017,Wang2019} and quantum dots \cite{Luxmoore2013,Arcari2014,Yalla2014,Sollner2015,Morrissey2009}. Overall, structured wQED provides a versatile route for programmable control of quantum-state transport, with direct implications for state transfer, quantum information routing, and collective emission engineering in integrated quantum networks \cite{Sunami2025}.

This paper is organized as follows. In Sec. \ref{sec.theoretical}, we introduce our system setup and the theoretical model of structured wQED with patterned coupling directionalities. In Sec. \ref{sec.excitaion}, we study the excitation dynamics for each structure and its properties either from the spectral analysis or the analytical solution in a smaller system size. We further reveal the spreading excitation dynamics in Sec. \ref{sec.spreading}, showing the excitation distribution behaviors in long-time relaxation dynamics. In Sec. \ref{sec.imperfection} we discuss the effect of nonguided mode to show the experimental feasibility in this model. Finally, we discuss and conclude in Sec. \ref{sec.discuss}.

%=========================
%========================

%%%%%%%%%%%%%%
\section{Theoretical Model} \label{sec.theoretical}
Controlling excitation flow in waveguide QED systems requires that each emitter can direct its emission into the left- or right-propagating modes of the waveguide. One practical approach is the double-waveguide configuration \cite{Yang2022}, where two parallel waveguides guide light in opposite directions [Fig. \ref{Fig1}(a)]. By changing the distance of each atom from the two waveguides, the coupling to one channel can be reduced while the coupling to the opposite channel becomes stronger, resulting in unequal decay rates $\gamma_{\mu}^{A}$ and $\gamma_{\mu}^{B}$, respectively. Another approach proposes to use a single waveguide \cite{Lee2025}, where the atoms are arranged in specific spatial patterns along the waveguide to control the excitation localization properties. This geometry changes the coupling strength through the local field distribution, but all atoms still share the same ratio between left- and right-propagating emissions, giving a uniform directionality across the array. 

Here we introduce a structured wQED model, where each emitter is assigned a local directionality that can vary site by site while the overall coupling strength remains uniform. This isolates the effect of spatially varying directionality patterns as the key resource for manipulating excitation dynamics. As illustrated in Fig. \ref{Fig1}(b), the structured wQED framework effectively maps the double-waveguide architecture onto an equivalent single-waveguide system, in which local directionality encodes the emission asymmetry otherwise determined by geometric coupling.

%========================
\begin{figure*}[t]
	\centering
	\includegraphics[width=0.97\textwidth]{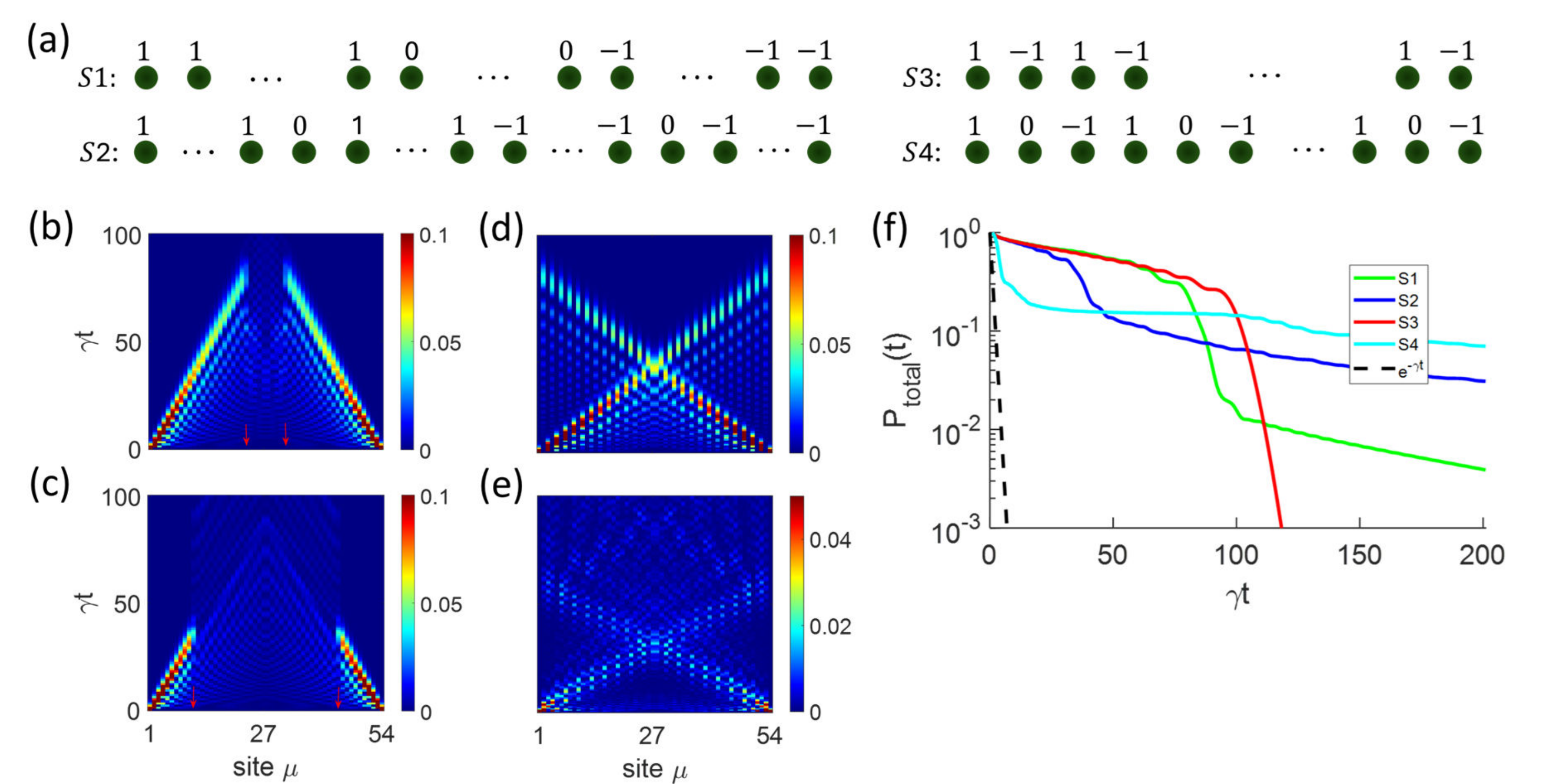}
	\caption{Excitation dynamics for four structured directionality configurations S1–S4.
		(a) Schematic representation of the four investigated configurations, labeled S1, S2, S3, and S4.
		(b–e) Corresponding excitation transport dynamics for each structure: (b) S1 – centering excitation, (c) S2 – wave-like excitation, (d) S3 – leap-frog excitation, and (e) S4 – dispersion excitation. The array consists of $N=54$ atoms separated by $\xi = \pi/2$. The red arrows in (b) and (c) indicate the sites where the directionality is changed to $D = 0$ accordingly. (f) Total population dynamics on a logarithmic scale. Solid lines indicate S1 (green), S2 (blue), S3 (red), and S4 (cyan). Black dashed line in (e) represent exponential decays $e^{-\gamma t}$, serving as references for comparing decay rates.}\label{Fig2}
\end{figure*}
%========================

With this model, we consider 1D arrays of two-level quantum emitters (the ground $|g\rangle$ and the excited state $|e\rangle$ as a spin-$1/2$ system) coupled to a nanophotonic waveguide \cite{Sheremet2023}, where photon-mediated spin-exchange interactions emerge among every pairwise atoms \cite{Pichler2015, Jen2025} due to the guided modes. The emitters are uniformly arranged by a distance $d$, where we quantify them in terms of dimensionless parameter $\xi\equiv k_sd$ with an atomic transition wavevector $k_s$. This distance is considered to be far enough that the near-field effect does not play an important role \cite{Kuraptsev2020}. Therefore, our theoretical model here closely follows the conventional waveguide QED system \cite{Sheremet2023} where the influence of short-range interaction can be negligible. The density matrix $\rho$ of $N$ atoms with nonreciprocal decay channels in an interaction picture evolves as ($\hbar=1$) \cite{Pichler2015},
\bea
\frac{d\rho}{dt}=-i[H_L +H_R,\rho] + \mathcal{L}_L[\rho] + \mathcal{L}_R[\rho],\label{rho}
\eea
where the coherent and dissipative system dynamics are determined by
\bea
H_{L(R)}=&&-\frac{i}{2}\sum_{\mu<(>)\nu}\sum_{\nu=1}^N \sqrt{\gamma_{L_\mu(R_\mu)}\gamma_{L_\nu(R_\nu)}}\nonumber\\
&&\times(e^{ik_s| r_\mu - r_\nu |}\sigma_\mu^\dagger\sigma_\nu - \rm{H.c.}),\label{H}
\eea
and
\bea
\mathcal{L}_{L(R)}[\rho]=&&-\frac{i}{2}\sum_{\mu,\nu = 1}^N \sqrt{\gamma_{L_\mu(R_\mu)}\gamma_{L_\nu(R_\nu)}} e^{\mp ik_s(r_\mu - r_\nu)} \nonumber\\
&&\times (\sigma_\mu^\dagger\sigma_\nu\rho+ \rho \sigma_\mu^\dagger\sigma_\nu - 2\sigma_\nu\rho\sigma_\mu^\dagger),\label{L}
\eea
with $\sigma_\mu^\dagger\equiv\vert e \rangle_\mu\langle g\vert$ and $\sigma_\mu = (\sigma_\mu^\dagger)^\dagger$. Here, $\gamma_{R_\mu(L_\mu)}$ denote the directional decay rates for emitter $\mu$ into the right (left) propagating modes, which in general may differ, i.e., $\gamma_{L_\mu} \neq \gamma_{R_\mu}$. The above effective density matrix equations are obtained under the Born–Markov approximation \cite{Lehmberg1970} by assuming an effectively 1D photonic reservoir \cite{Tudela2013}, where the guided modes mediate long-range and all-to-all dipole–dipole interactions \cite{Mitsch2014, Sheremet2023}. 

The degree of nonreciprocity in photon exchange is quantified by a local directionality factor for each emitter, defined as $D_\mu = (\gamma_{R_\mu} - \gamma_{L_\mu})/\gamma$ \cite{Mitsch2014}, where $\gamma = \gamma_{R_\mu} + \gamma_{L_\mu} \equiv 2|dq(\omega)/d\omega|_{\omega = \omega_{eg}} g_{k_s}^2 L$ \cite{Tudela2013}. Here $|dq(\omega)/d\omega|$ represents the inverse group velocity at resonance, $g_{k_s}$ is the atom–waveguide coupling strength, and $L$ is the quantization length. The set $\{D_\mu\}$ defines a spatial directionality pattern composed of discrete values $D_\mu \in \{+1,0,-1\}$: sites with $D_\mu = +1$ ($-1$) couple exclusively to the right- (left-) propagating mode, while $D_\mu = 0$ represents a locally reciprocal site acting as a source or scattering node. To generalize the model, we introduce a chirality factor $\eta \in [0,1]$ such that $D_\mu \rightarrow \eta D_\mu$, where $\eta=1$ indicates the directionality pattern to contain fully cascaded (unidirectional) and reciprocal regimes, while a smaller $\eta$ describes a nonreciprocal regime with a fractional chirality.

We initialize the system with a single atomic excitation, and hence, the system evolution can be well described by the single-excitation subspace spanned by the basis states $\vert \psi_\mu \rangle = \vert e \rangle_\mu \vert g \rangle^{\otimes(N-1)}$. Within this subspace, the general dynamical equations of a density matrix can be reduced to non-Hermitian Schr\"{o}dinger equation with an effective Hamiltonian $H_{\rm eff}$ \cite{Pichler2015,Yang2022},   
\bea
i\frac{\partial}{\partial t}\vert \Psi(t)\rangle = H_{\rm eff}\vert \Psi(t)\rangle, \label{Psi}
\eea
where
\bea
H_{\rm eff} = &&-\frac{i}{2}\sum_{\mu<\nu}^N \sqrt{\gamma_{L_\mu}\gamma_{L_\nu}}e^{ik_s| r_\mu - r_\nu |}\sigma_\mu^\dagger\sigma_\nu \notag\\
&&-\frac{i}{2}\sum_{\mu>\nu}^N \sqrt{\gamma_{R_\mu}\gamma_{R_\nu}}e^{ik_s| r_\mu - r_\nu |}\sigma_\mu^\dagger\sigma_\nu \notag\\
&&- \frac{i}{2}\sum_{\mu=1}^N\gamma\sigma_\mu^\dagger\sigma_\mu.\label{Heff}
\eea
Here, the state of the system is $\vert \Psi(t)\rangle =\sum_{\mu = 1}^N a_\mu (t)\vert \psi_\mu\rangle$ with the probability amplitudes $a_\mu (t)$. Taking into account of the local directionality for each emitter, we obtain the coupled equations of the system as
\bea
\dot{a}_\mu(t) = &&-\sum_{\nu>\mu}^N \sqrt{\gamma_{L_\mu}\gamma_{L_\nu}}e^{-ik_s(r_\nu-r_\mu)}a_\nu(t)-\frac{\gamma}{2}a_\mu(t)\nonumber\\
&&-\sum_{\nu<\mu}^N\sqrt{\gamma_{R_\mu}\gamma_{R_\nu}} e^{-ik_s(r_\mu-r_\nu)}a_\nu(t).\label{eqdiffamplitude}
\eea
We can analyze the excitation dynamics by solving the above equations for different configurations and identify distinct propagation behaviors depending on the pattern of $D_\mu$ in the next section. 

%%%%%%%%%%%%%%
\section{Excitation Dynamics}\label{sec.excitaion}
Here we study excitation transport in one-dimensional arrays of quantum emitters arranged in four representative directionality patterns—S1, S2, S3, and S4—as illustrated in Fig. \ref{Fig2}(a). Each structure can be defined by site-dependent directionality $D_\mu$ with $\eta = 1$. Specifically, they are designed to capture distinct combinations of symmetry, the presence of reciprocal sites ($D_\mu = 0$), and alternation of chiral couplings ($D_\mu = \pm 1$), which are key ingredients that govern excitation flow in structured wQED.	All patterns preserve the mirror symmetry but differ in how reciprocal sites are embedded, leading to qualitatively different interference mechanisms. S1 considers three zones where reciprocal sites are placed and sandwiched between two groups of chiral couplings, while S2 is embedded with reciprocal sites within respective zones with chiral couplings. S3 represents a fully alternating pattern that maximizes nonlocal chiral transfer, while S4 introduces a periodic mixing of unidirectional and reciprocal couplings within a unit cell.	Together, these structures form a compact set of representative patterns that illustrate how different directionality pattern give rise to distinct transport behaviors and provide design guidelines for constructing other structured configurations. 
	
We note that the structure of S3 is equivalent to two effectively decoupled atomic arrays evolving independently with respective chiral excitation transport without interferences or population exchanges among the subsystems with different chiralities. Therefore, S3 serves as a representative structure wQED, where a directional transport can be achieved in a controllable way. This is well understood in a cascaded platform \cite{Carmichael1993, Stannigel2012, Jen2020_collective}, but in Sec. \ref{sec.spreading} we will release this condition of $\eta=1$, and the case with $\eta<1$ will lead to more centralized population distributions due to the influence of finite reciprocal couplings and excitation interferences, where a picture of decoupled atomic arrays is no longer valid.

S1 follows $(1,1,\cdots,1,0,\cdots,0,-1,\cdots,-1)$, forming mirror-symmetric right and left regions that funnel excitation from both edges toward the central reciprocal segment. 
S2 adopts $(1,\cdots,1,0,1,\cdots,1,-1,\cdots,-1,0,-1,\cdots,-1)$, a mirror-symmetric pattern with inserted reciprocal sites ($D_\mu = 0$) that locally re-couple the left and right propagating channels, allowing excitation to be redistributed from these sites. S3 alternates $(+1,-1,+1,-1,\cdots)$, dividing the array into two interlaced sublattices that exchange excitation nonlocally. S4 repeats $(+1,0,-1,+1,0,-1,\cdots)$, combining unidirectional and reciprocal couplings within each three-site unit cell. Unless stated otherwise, the array contains $N=54$ emitters separated by a uniform interatomic spacing $\xi = \pi/2$. 

In Fig. \ref{Fig2}(b-e) we observe distinct excitation dynamics associated with each structure -- centering, wave-like, leap-frog, and dispersion excitation -- reflecting the interplay between local directionality and the collective photon-mediated coupling. As shown in Fig. \ref{Fig2}(f), although all structures eventually exhibit population decay over long times, their total excitation populations decrease significantly slower than a decay of atom in free space $e^{-\gamma t}$, which is a hallmark of collective subradiance in 1D emitter–waveguide coupled systems \cite{Zhang2019,Albrecht2019,Pennetta2022, Jen2025}. This subradiant behavior arises from photon-mediated dipole-dipole interactions, leading to a suppression of atomic radiations under structured directionalities. 

We note that the sudden drops in the total population observed in Fig. \ref{Fig2}(f) directly correlated with specific dynamical mechanisms underlying each structured configuration. In the fully alternating leap-frog configuration S3, the sharp population drop originates from boundary-limited chiral propagation: once the excitation reaches both ends of the array, further transport is blocked and rapid radiative decay occurs. In contrast, configurations involving reciprocal sites (S1, S2, and S4) exhibit milder population drops that are associated with excitation transport across reciprocal region. In these cases, the long-time dynamics is subsequently dominated by interference and beating between long-lived subradiant collective modes, resulting in a smoother decay profile.

We further analyze the excitation dynamics of structures S1, S2, and S4 through spectral analysis, while that of S3 is examined analytically using a small-size model, since its fully chiral pattern ($D = \pm1$) permits a closed-form solution, unlike the other configurations that involve reciprocal nodes ($D = 0$). For the spectral analysis, we choose $\eta=0.999$ to approximate the fully cascaded regime ($D_\mu=\pm 1$) while preserving numerical stability, since $\eta=1$ would induce exceptional points that are not diagonalizable. The effective non-Hermitian Hamiltonian $H_{\mathrm{eff}}$ governs both the coherent and dissipative couplings in the system, and its complex eigenvalues encode the collective frequencies and decay rates. The right and left eigenstates satisfy \cite{Ashida2020}
\bea
H_{\mathrm{eff}}\ket{\phi_n^{R}}=\lambda_n\ket{\phi_n^{R}}, \qquad
\bra{\phi_n^{L}}H_{\mathrm{eff}}=\lambda_n\bra{\phi_n^{L}},
\eea
and obey the biorthogonality condition $\braket{\phi_m^{L}|\phi_n^{R}}=\delta_{mn}$, where the eigenvalue $\lambda_n=\omega_n-i\gamma_n$ contains the collective frequency shift $\omega_n$ and radiative decay rate $\gamma_n$. The time evolution of an arbitrary initial excitation $\ket{\Psi(0)}$ can thus be expanded as
\bea
\ket{\Psi(t)}=\sum_n \alpha_n\, e^{-i\lambda_n t}\ket{\phi_n^{R}}, \qquad
\alpha_n=\braket{\phi_n^{L}|\Psi(0)}.\label{eqspektral}
\label{spectral_solution}
\eea
Each eigenmode oscillates at frequency $\omega_n$ and decays at rate $\gamma_n$, while their interference governs the collective excitation dynamics.

To capture the late-time relaxation behavior, we selectively excite a subset of eigenmodes that dominantly contribute to the long-time dynamics under our considered state initializations quenched at the system boundaries. This subset $S$ represents the collective modes whose interference reproduces the characteristic spatial–temporal patterns observed in the system. The resulting state can be written as a superposition of these modes,
\bea
\ket{\Psi(0)}=\sum_{n\in S} a_n\,\ket{\phi_n^{R}}, \qquad
\ket{\Psi(t)}=\sum_{n\in S} a_n\,e^{-i\lambda_n t}\ket{\phi_n^{R}},
\label{superposition}
\eea
which highlights the interference among collective eigenmodes and reproduces the spatial–temporal excitation dynamics shown in Figs. \ref{Fig3}–\ref{Fig5}. Each eigenvector can be written in the site basis as
\bea
\ket{\phi_n^R} = \sum_{\mu} v_{\mu}^{(n)} \ket {\psi_\mu},
\eea
where $v_{\mu}^{(n)}$ is, in general, a complex-valued amplitude. So, the phase is defined as
\bea
	\varphi_{\mu}^{(n)} = \arg\!\left(v_{\mu}^{(n)}\right)
	= \tan^{-1}\!\left(\frac{\operatorname{Im}[v_{\mu}^{(n)}]}{\operatorname{Re}[v_{\mu}^{(n)}]}\right),
\eea
measured relative to a chosen reference site.

In below, we analyze the dynamical behaviors and spectral properties of each structured configuration, S1–S4. For S1, S2, and S4, the excitation dynamics are examined through spectral decomposition of the effective non-Hermitian Hamiltonian, whereas S3 is discussed analytically using a simplified four-atoms model. This section clarifies how the dominant eigenmodes and their interferences manifest the distinct transport features observed in Fig.\ref{Fig2} as well as how the excitation can be transported if initialized from the both edges. We further analyze the emergence of dark modes in the S1 configuration as an illustrative example, and explicitly demonstrate their decoherence-free nature.

\subsection{S1: Centering excitation}
For the S1 configuration, the directionality sequence divides the array into two mirror-symmetric chiral domains separated by several central emitters with reciprocal couplings. A majority of atoms in the left or right parts couple only to the right-propagating or the left-propagating modes, respectively. Consequently, an excitation initially localized at either edge propagates inward from both sides, converging toward the middle where the reciprocal couplings allow the interference between counter-propagating components. This interference gives rise to the centering dynamics shown in Fig. \ref{Fig2}(b), characterized by a continuous accumulation of population on the central area of the array and a relatively slow overall decay compared with other structures.

%========================
\begin{figure}[t]
	\centering
	\includegraphics[width=0.49\textwidth]{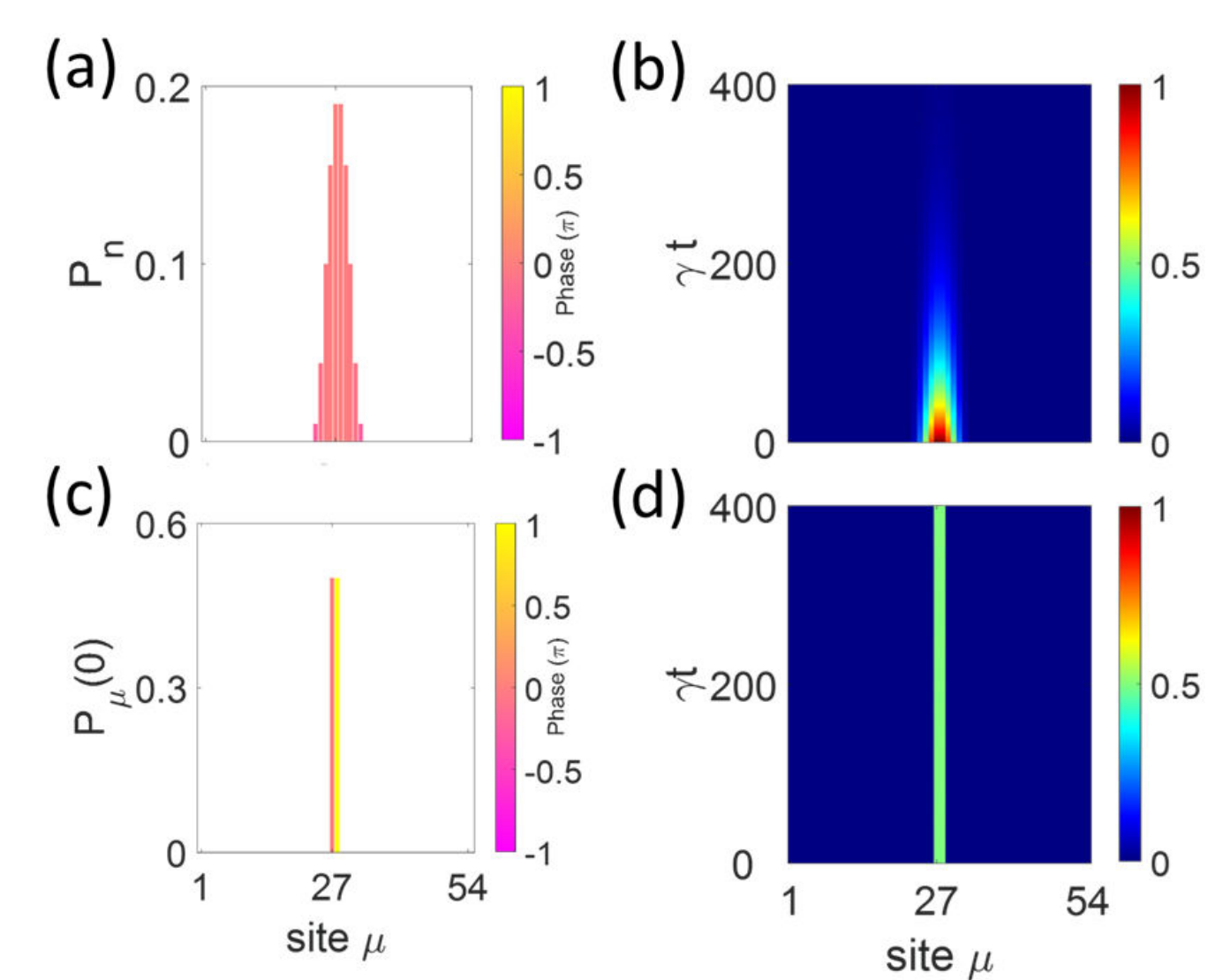}
	\caption{A middle-localized subradiant mode found in structure S1, which the $D_\mu = 0$ placed in sites 23 to 32, with $\eta = 0.999$. (a) The population profile of the most subradiant mode ($\lambda_1 = -0.5250\gamma - i0.00581\gamma$). (b) The time evolution of the system initially quenched at this subradiant mode. Other parameters, including the interatomic spacing $\xi$ and the number of atoms $N$, are the same as in Fig. \ref{Fig2}(b). (c) An example for one of the degenerate dark states $\vert \rm Dark\rangle = (\vert \psi_{27}\rangle - \vert \psi_{28}\rangle)/\sqrt{2}$ and (d) its time evolution by choosing Bragg spacing, $\xi = 2\pi$.}\label{Fig3}
\end{figure}
%========================

The symmetric bidirectional flow effectively traps parts of the excitation in the middle region through destructive interference among the central areas. This mechanism tangentially mimics the excitation trapping effect \cite{Handayana2024} as well as subradiance-protected transport \cite{Needham2019}, and behaves like a self-formed cavity where the central atoms experience enhanced reabsorption before eventually decaying through the dissipation mechanism, as shown in Fig. \ref{Fig2}(f). 

By spectral analysis in Eq. \ref{eqspektral}, we demonstrate that excitation trapping mainly contributed by a special subradiant mode, see Fig. \ref{Fig3}(a,b). This subradiant mode is localized in the middle area where $D_\mu = 0$, having a small decay rate that ensures long-lived excitation confinement of the trapped excitation. However, this mode underlies the weak residual excitation near the center in Fig. \ref{Fig2}(b), which becomes dominant only at late times after faster-decaying components vanish. Furthermore, if we choose $\xi = \pi$ or $2\pi$, the atoms would form a Bragg-spaced array, enhances the coherence of photon-mediated interactions, and the dark modes would emerge in the system. Therefore, the dominant subradiant mode in Fig. \ref{Fig3} resembles some of the features of these dark modes, which can persist at long time and become concentrated eventually.

To investigate the emergence of dark states in structure S1, we consider a five-atoms system as an example. The directionality pattern is chosen as (+$\eta$, 0, 0 , 0, -$\eta$) corresponding to two outer unidirectional emitters enclosing three reciprocal sites, and the interatomic spacing is fixed at the Bragg condition $\xi = 2\pi$. Under these conditions, the effective non-Hermitian Hamiltonian in the single-excitation manifold of Eq. (\ref{Heff}) takes the matrix form
\bea
H_{\mathrm{eff}} = -i\,\frac{\gamma}{2}
\begin{pmatrix}
	1 & \Lambda^- & \Lambda^- & \Lambda^- & \Lambda^+\Lambda^- \\
	\Lambda^+ & 1 & 1 & 1 & \Lambda^+  \\
	\Lambda^+ & 1 & 1 & 1 & \Lambda^+  \\
	\Lambda^+ & 1 & 1 & 1 & \Lambda^+  \\
	\Lambda^+\Lambda^- & \Lambda^- &\Lambda^- & \Lambda^- & 1
\end{pmatrix}
\eea
with $\Lambda^+ = \sqrt{1+\eta}$ and $\Lambda^- = \sqrt{1-\eta}$. One can notice that the middle three columns are identical, which suggest that we can form dark dimer states by choosing any two of the atoms on these three reciprocal sites \cite{Poshakinskiy2021}. For these three atoms, we can form two linear independent dark dimer states with the wave function $\vert \text{Dark}_1\rangle = \left(\vert \psi_2 \rangle-\vert\psi_3\rangle\right)/\sqrt{2}$ and $\vert \text{Dark}_2\rangle = \left(\vert \psi_3 \rangle -\vert\psi_4\rangle\right)/\sqrt{2}$. These dark state do not dissipate since $H_{\rm eff}\vert \text{Dark} \rangle = 0$. More generally, for an S1 configuration containing $n$ reciprocal sites at Bragg spacing $\xi = \pi, 2\pi$, the dimension of the dark-state manifold is $n-1$, reflecting the rank deficiency of the collective coupling matrix. The decoherence-free nature for one of the degenerate dark states $\vert \rm Dark\rangle = (\vert \psi_{27}\rangle - \vert \psi_{28}\rangle)/\sqrt{2}$ is explicitly demonstrated in Fig. \ref{Fig3}(c,d) for 54 atoms.

\subsection{S2: Wave-like excitation}
In the S2 configuration, we insert two reciprocal sites within the chiral regions, $D_\mu=\pm 1$, respectively. The local reciprocal sites with $D_\mu = 0$ act as sources that radiate excitation into adjacent chiral regions, while the sites with $D_\mu = \pm1$ serve as transport channels for unidirectional propagations. The reciprocal sites act as sources of radiations, while $D\mu=\pm 1$ as transport channels for unidirectional propagation. This alternating combination allows excitation to be redistributed from the reciprocal sites, creating a sequential population transfer that closely resembles a traveling wave. As shown in Fig. \ref{Fig2}(c), an excitation initiated at both edges propagates successively through the array with a well-defined phase front, leaving behind a slowly decaying intensity trail.

%========================
\begin{figure}[t]
	\centering
	\includegraphics[width=0.49\textwidth]{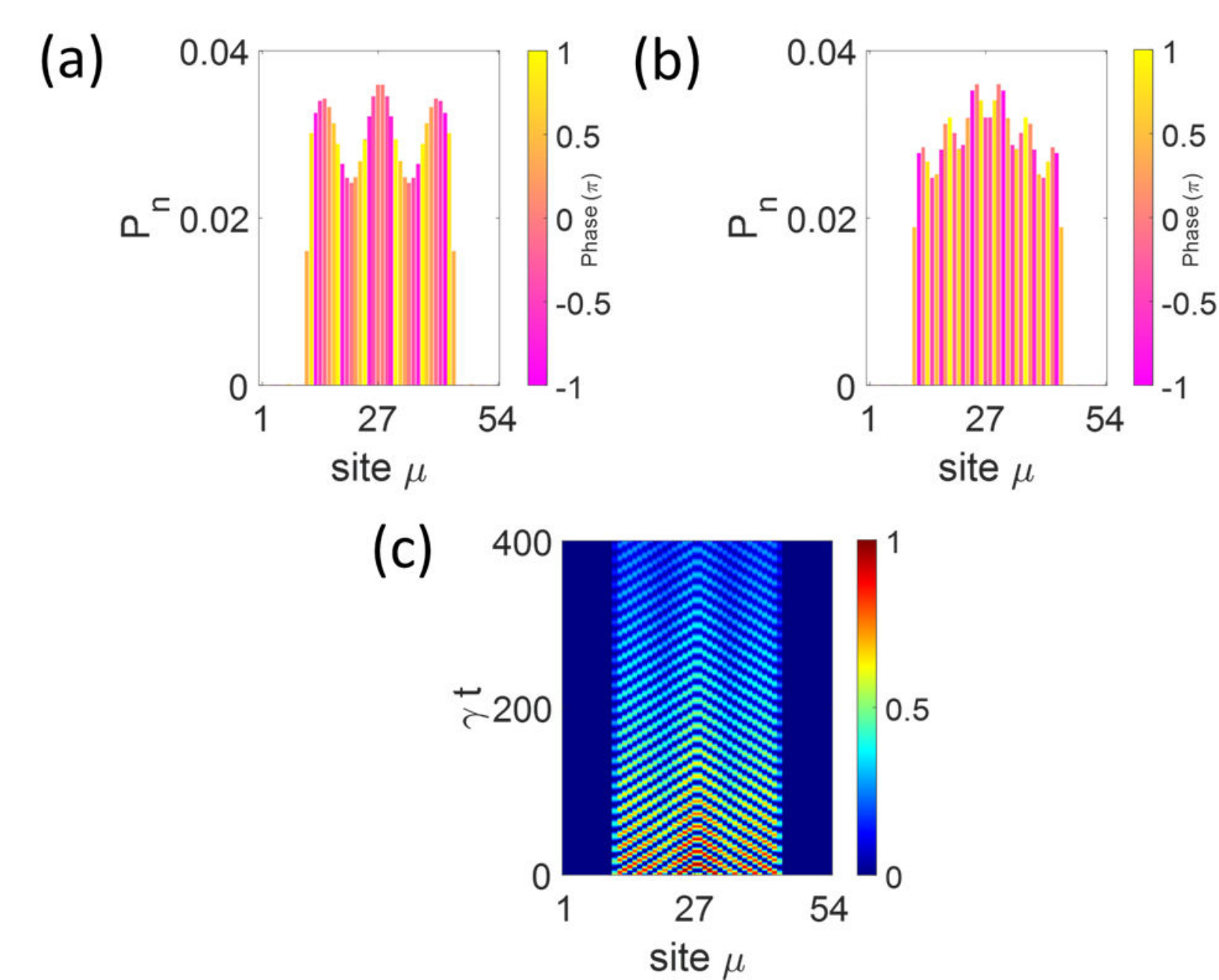}
	\caption{Two middle-localized subradiant modes and their beating dynamics in the S2 configuration, with $D_\mu = 0$, placed at sites 11 and 44. Population profile of a subradiant mode with frequency shifts (a) $\omega_5 = -0.126\gamma$ and (b) $\omega_{12} = 0.291\gamma$. (c) Time evolution of a superposition of these two modes, showing a clear beating pattern with a frequency $\omega_{\mathrm{beat}} = \omega_{12} - \omega_{5}$ that reproduces the wave-like excitation observed in Fig. \ref{Fig2}(c). Other system parameters are the same as in Fig. \ref{Fig3}.}\label{Fig4}
\end{figure}
%========================

In contrast to the centering configuration S1, where excitation energy accumulates near the center, the S2 structure sustains a continuous directional flow mediated by these embedded reciprocal nodes, giving rise to a coherent wave-like pattern across the system. The reciprocal sites interrupt the unidirectional propagation from both sides, creating points of constructive interference between forward and backward excitations. This mechanism effectively stabilizes the transport, resulting in a slow decay excitation population $P_{total}(t)$, as shown in Fig. \ref{Fig2}(f). This coherent wave-front is akin to interference-stabilized excitation transfer via many-body dark and subradiant channels in uniform arrays \cite{Chen2025}, here realized geometrically by embedding $D_\mu=0$ nodes.

By decomposing the excitation dynamics into the eigenmodes of $H_{\mathrm{eff}}$ in Eq. \ref{eqspektral}, we find that the long-time behavior of S2 can be constructed by a few subradiant modes [Fig. \ref{Fig4}(a,b)]. These modes are spatially localized between the two reciprocal sites ($D_\mu=0$) and exhibit population profiles with a triangular envelope. The localization distributions showcase why the excitation remains confined in the central region, while the triangular envelope determines the shape of the propagating wave front. The characteristic wave-like dynamics arises from the interference between subradiant modes with slightly different energy shifts, which generates a temporal beating pattern. An example of this beating, obtained from the superposition of two dominant subradiant modes, is shown in Fig. \ref{Fig4}(c); the period of the oscillating fronts corresponds to the beating frequency $\omega_{\mathrm{beat}} = \omega_{12} - \omega_{5}$. In general, if $n$ subradiant modes contribute, the dynamics contains ${n \choose 2}$ distinct beating frequencies that generate multiple interference periods in the excitation patterns.

\subsection{S3: Leap-frog excitation}
In the S3 configuration, an alternating directionality pattern $(+1,-1,+1,-1,\cdots)$ divides the array into two interlaced sublattices that couple exclusively to opposite propagation directions. This arrangement enables nonlocal excitation transfer between distant sites, creating a “leap-frog” dynamic where population jumps between atoms separated by one site. As shown in Fig. \ref{Fig2}(d), an excitation initially localized at one edge hops between alternating sites. This behavior arises because each atom emits into a mode propagating away from its nearest neighbors, leading to a nonlocal coupling that bypasses intermediate sites. The resulting dynamics are characterized by rapid oscillations between the two sublattices, with minimal occupation of the sites in between. 

To illustrate this behavior, we can analytically solve the dynamics for a four-atom system in the S3 configuration \cite{Jen2020_collective}. From Eq. \eqref{eqdiffamplitude}, the coupled equations for the amplitudes are
\bea
\dot a_1(t)&=&-\frac{\gamma}{2}a_1(t), \notag\\
\dot a_2(t)&=&-\frac{\gamma}{2}a_2(t)-\gamma e^{-i2\xi} a_4(t), \notag\\
\dot a_3(t)&=&-\frac{\gamma}{2}a_3(t)-\gamma e^{-i2\xi} a_1(t),\notag\\
\dot a_4(t)&=&-\frac{\gamma}{2}a_4(t).
\eea
Solving these equations gives the time evolution of the amplitudes
\bea
a_1(t)&=&a_1(0)\,e^{-\frac{\gamma}{2}t}, \notag\\
a_2(t)&=&\left[a_2(0)-\gamma\,t\,e^{-i2\xi}\,a_4(0)\right]e^{-\frac{\gamma}{2}t}, \notag\\
a_3(t)&=&\left[a_3(0)-\gamma\,t\,e^{-i2\xi}\,a_1(0)\right]e^{-\frac{\gamma}{2}t}, \notag\\
a_4(t)&=&a_4(0) e^{-\frac{\gamma}{2}t}.
\eea
This solution precisely demonstrates the leap-frog mechanism. The amplitude $a_1(t)$ and $a_4(t)$ evolves independently, completely decoupled from the other two. The dynamics of the atoms 2 and 3, however, are directly driven by the atoms 4 and 1, respectively. This shows that the photon-mediated interaction effectively bypasses the intermediate site, creating a nonlocal transfer of excitation between the odd or even sites.

Interestingly, the excitation transport in the S3 configuration is ultimately limited by the system boundaries. Once the propagating excitation reaches the array edges, it can no longer transfer further and undergoes rapid radiative decays. This explains the sharp drop of the total population around $\gamma t = 100$ in Fig. \ref{Fig2}(f), which marks the moment when the excitation propagation simultaneously touches both boundaries when an atomic excitation is  initialized simultaneously at the edges. Consequently, the transport lifetime in this configuration scales with the array length: larger systems allow longer propagation before the excitation is lost, thereby extending the overall survival time of the collective state. 

\subsection{S4: Dispersion excitation}
The S4 configuration follows a repeating three-site pattern $(+1,0,-1,+1,0,-1,\cdots)$,
where right-coupled, reciprocal, and left-coupled emitters are interleaved periodically.
This structure mixes opposite propagation channels within each unit cell, creating a strongly modulated local environment that promotes interference between forward and backward of atomic excitations. As a result, an excitation initialized at each edge forms a disperse excitation, as shown in Fig. \ref{Fig2}(e). The population becomes broadly distributed across the array, and the spatial profile fluctuates dynamically due to the competing couplings among neighboring sites.

%========================
\begin{figure}[b]
	\centering
	\includegraphics[width=0.49\textwidth]{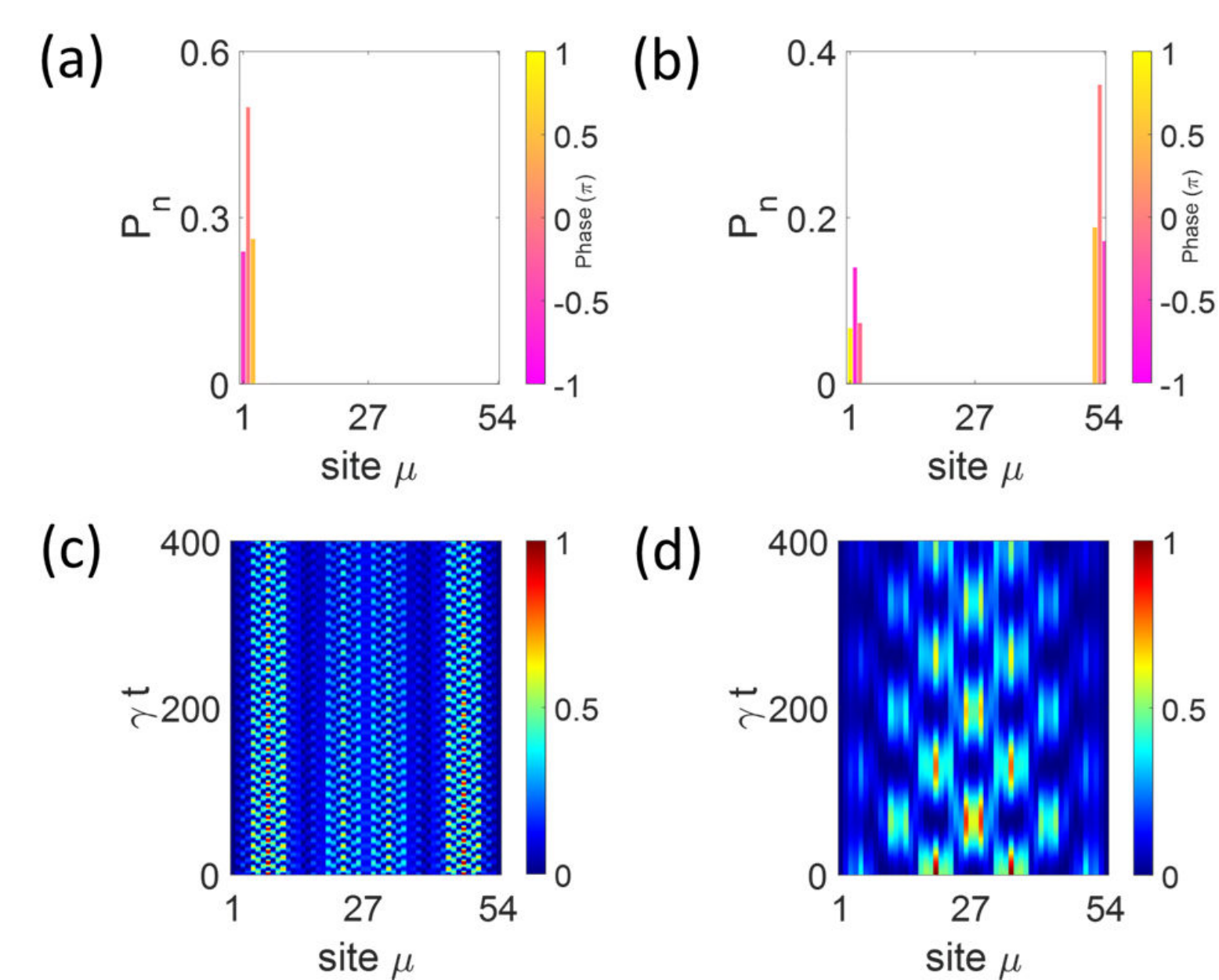}
	\caption{Two edge modes and distinct beating dynamics in the S4 configuration. (a,b) Population profiles of two nearly degenerate eigenmodes localized at the first and last unit cells, corresponding to edge states arising from the broken translational symmetry of the directionality pattern. (c,d) Time evolution of the superposition of two subradiant modes with a high beat frequency (c) $\omega_{\mathrm{beat}} = 0.467\gamma$, forming a herringbone-like interference pattern and (d) $\omega_{\mathrm{beat}} = 0.048\gamma$, contributing to the global dispersive behavior across the array. Other system parameters are the same as in Fig. \ref{Fig3}.}\label{Fig5}
\end{figure}
%========================

Because of this strong mode mixing, the S4 configuration exhibits slower decay behavior, as evident in Fig. \ref{Fig2}(f). The interplay between local cascade and reciprocal nodes leads to partial reabsorption and irregular interference that sustain residual excitations for longer times, even as the system becomes spatially disperse. This structure shows continuous fluctuation throughout the array and aligns with the emergence of spatially extended collective modes in waveguide-coupled emitter arrays \cite{Masson2020}. 

The spectral properties of S4 explain several unique features of its dynamics. First, the earlier decay of the total population in Fig. \ref{Fig2}(f) originates from the presence of edge-localized modes [Fig. \ref{Fig5}(a,b)]. These states emerge due to the broken translational symmetry \cite{Lodahl2017}, which supports two nearly degenerate eigenmodes confined to the first and last unit cells. Because these edge states possess relatively large decay rates, an excitation initialized near the boundaries effectively excites these lossy modes, leading to rapid population loss localized at the edges. 

In addition to the edge effects, S4 exhibits two qualitatively different interference dynamics governed by the beating between subradiant modes. A high-frequency beating between modes with large frequency detuning produces a herringbone-like interference pattern [Fig.\ref{Fig5}(c)], characterized by rapid oscillations and fine spatial fringes. In contrast, low-frequency beatings between closely spaced modes [Fig.\ref{Fig5}(d)] lead to slower, large-scale oscillations and smooth population redistribution across multiple unit cells. The longer period associated with low-frequency interference marks a distinction between adjacent unit cells clearly visible, in contrast to the fast oscillations produced by high-frequency beatings. When the interatomic spacing is tuned to $\xi = \pi$ or $2\pi$, dark modes analogous to those in S1 reappear. These modes extend across all unit cells except the edges and form a decoherence-free subspace, allowing any excitation initialized away from the boundaries to remain trapped within its local unit cell indefinitely.

\subsection{Lifetime and population of subradiant modes}
Here we note that the lifetime of the dominant subradiant eigenmodes generally increases with the number of atoms, as collective destructive interference in the radiative channels becomes more effective in larger arrays. This qualitative scaling behavior is a well-known feature of waveguide QED and related collective-emission systems, where increasing system size promotes the suppression of radiative decay. Importantly, this trend remains qualitatively robust with respect to the interatomic spacing, provided that the spacing is chosen near interference-favorable conditions, such as the Bragg-like regimes discussed throughout this section (e.g., $\xi \approx \pi$ or $2\pi$).

Within the single-excitation manifold, the specific subradiant modes that become populated during the dynamics depend on the chosen initial condition. The time evolution can be understood by decomposing the initial state into the collective eigenmodes of the effective non-Hermitian Hamiltonian: modes with a significant overlap with the initial excitation are preferentially populated and thus contribute to the subsequent dynamics. In the long-time limit, faster-decaying modes are suppressed, and the population becomes dominated by those subradiant modes that are both long-lived and efficiently populated from the initial state. As a result, the modes governing long-time trapping are not necessarily the most subradiant ones available in the full spectrum, but rather the subset that combines small decay rates with appreciable initial overlap.

As a final remark, the effectiveness of a given directionality pattern in storing excitations at long times is primarily determined by the presence and spatial distribution of reciprocal sites ($D_\mu = 0$), which enable collective destructive interference between left- and right-propagating emission channels. Directionality patterns that contain extended reciprocal regions, or reciprocal sites separated over sufficiently large distances, favor the formation of collective modes with well-defined phase relations, leading to strong suppression of radiative decay and efficient long-time population trapping. In contrast, configurations with few or isolated reciprocal sites provide fewer opportunities for such interference to build up. In these cases, excitation more readily couples to unidirectional transport channels and boundary-loss mechanisms, populating modes with larger decay rates and resulting in weaker long-time storage, as shown by the abrupt population drop observed for the fully chiral leap-frog configuration S3 in Fig. \ref{Fig2}(f).

%%%%%%%%%%%%%%
\section{Spreading Characteristics of Excitation Dynamics}\label{sec.spreading}
Furthermore, to quantify the directional spreading of excitation, we evaluate the variance of the site populations, which can be defined as
\bea
s(t) = \sqrt{\sum_{j=1}^N\left(x_j-x_{cm}(t)\right)^2p_j(t)},
\eea
where $x_j=2\frac{j-1}{N-1}-1$ maps site positions within $[-1,1]$, and designates the center of the array as the site 0. $x_{cm}(t)=\sum_j x_jp_j(t)$ is the "center of mass" of the excitation distribution at time $t$, and $p_j(t)=\frac{n_j(t)}{\sum_k n_k(t)}$ is the normalized population with $n_j \equiv \langle\sigma^\dagger_j \sigma_j\rangle$. 

\begin{figure}[t]
	\centering
	\includegraphics[width=0.48\textwidth]{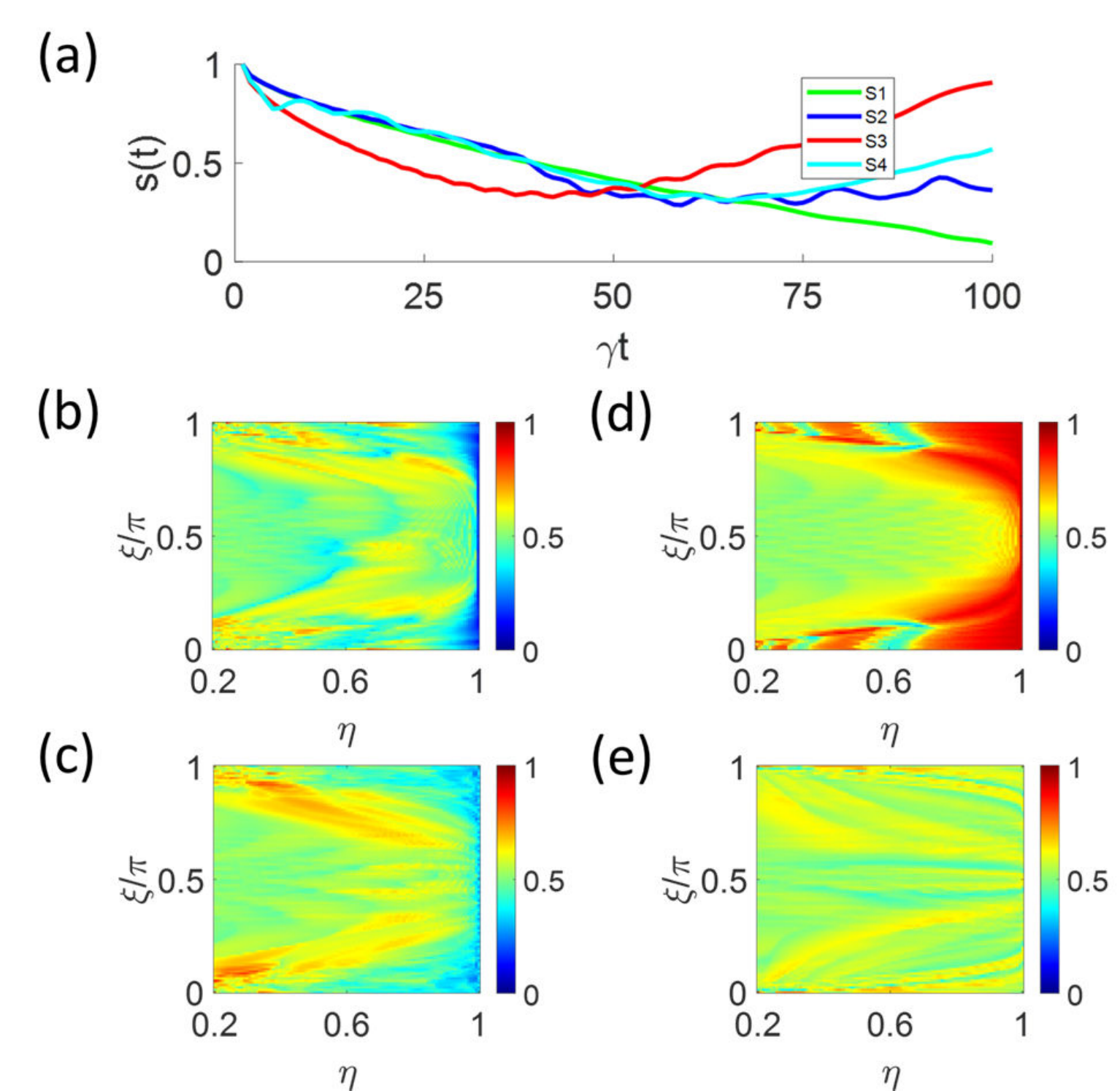}
	\caption{(a) Time evolution of the spreading population as variance, $s(t)$ for $\eta = 1$. (b-e) Spread $s_\text{st}$ in steady condition ($P_{total}\leq 10\%$) as functions of the directionality factor $\eta$ and inter particle spacing $\xi$ for (b) S1, (c) S2, (d) S3, and (e) S4. The system size is the same as in Fig. \ref{Fig2}.}\label{Fig6}
\end{figure}

From Fig. \ref{Fig6}(a), the variance dynamics $s(t)$ reveal distinct spreading characteristics for each structured configurations. The S1 configuration shows a monotonic decrease of $s(t)$, indicating that the excitation becomes increasingly concentrated toward the center as time evolves—consistent with the centering behavior observed in Fig. \ref{Fig2}(b). For S2, $s(t)$ initially decreases and then saturates around $s(t) = 0.4$, reflecting the formation of a stable distribution concentrated within the region formed by the reciprocal sites $(D_\mu = 0)$ that act as sources of radiations. This produces the wave-like pattern seen in Fig. \ref{Fig2}(c). In contrast, the S3 configuration exhibits a nonmonotonic trend: $s(t)$ first decreases and then rises toward unity, signifying a crossing event where the excitation hops from one sublattice to the other, consistent with the leap-frog excitation dynamics illustrated in Fig. \ref{Fig2}(d). Finally, S4 shows a slight initial reduction of $s(t)$—associated with the short-range transport of excitation from the edge toward the interior—but subsequently stabilizes around $s(t) = 0.7$. This plateau indicates persistent dispersion caused by strong mode mixing within its repeating three-site pattern, as reflected in Fig. \ref{Fig2}(e).

%========================
%========================

To characterize the long-time excitation distribution, we evaluate the spread $s_\text{st} = s(t)\vert_{P_{total}(t)\approx 10\%}$ corresponding to the condition when the total population has reached around $10\%$. This criterion provides a robust measure of the intrinsic transport tendency, independent of transient oscillations at early times. Figures \ref{Fig6}(b-e) presents the quasi-steady-state spread $s_\text{st}$ as a function of the interatomic spacing $\xi$ and the directionality factor $\eta$ for the four structured configurations S1-S4, respectively. In general, for weak chirality (small $\eta$), the spreads are broad and display irregular patterns with respect to $\xi$, reflecting the competition between left- and right-propagating modes. As the chirality increases, more distinct features emerge. For instance, the centering structure S1 in Fig. \ref{Fig6}(b), shows a pronounced minimum of $s_\text{st}$ near high $\eta$, indicating strong concentrated of excitation toward the center. A similar but broader distribution trend appears in S2, as shown in Fig. \ref{Fig6}(c), where the wave-like configuration maintains a moderate spread across a wide range of $\eta$, suggesting stable directional transport with partial confinement. In contrast, S3 exhibits high variations when $\eta$ is relatively high, which shows a significant contribution of leap-frog propagation at late time, as shown in Fig. \ref{Fig6}(d). Since it has clear excitation propagation symmetry, the $s_\text{st}$ also shows symmetric behavior with respect to $\xi$ for particular $\eta$. Lastly, S4 shows relatively uniform values of $s_\text{st}$ around 0.6–0.8, consistent with the dispersion-dominated propagation in which the excitation remains delocalized but not completely diffusive.

Overall, this analysis reveals that increasing directionality ($\eta \rightarrow 1$) generally enhances localization in symmetric or partially reciprocal arrays (S1 and S2), which resonate with concentrated subradiant patterns \cite{Yang2022}, whereas alternating or mixed-chirality structures (S3 and S4) retain broader spreads due to strong mode mixing \cite{Mirza2017, Fayard2021}. This demonstrates that the local engineering of directionality patterns provides an effective way to tune the degree of excitation confinement and the emergence of subradiant transport regimes in structured waveguide QED systems.

%%%%%%%%%%%%%%
\section{Imperfections and Experimental feasibility}\label{sec.imperfection}
\begin{figure}[t]
	\centering
	\includegraphics[width=0.47\textwidth]{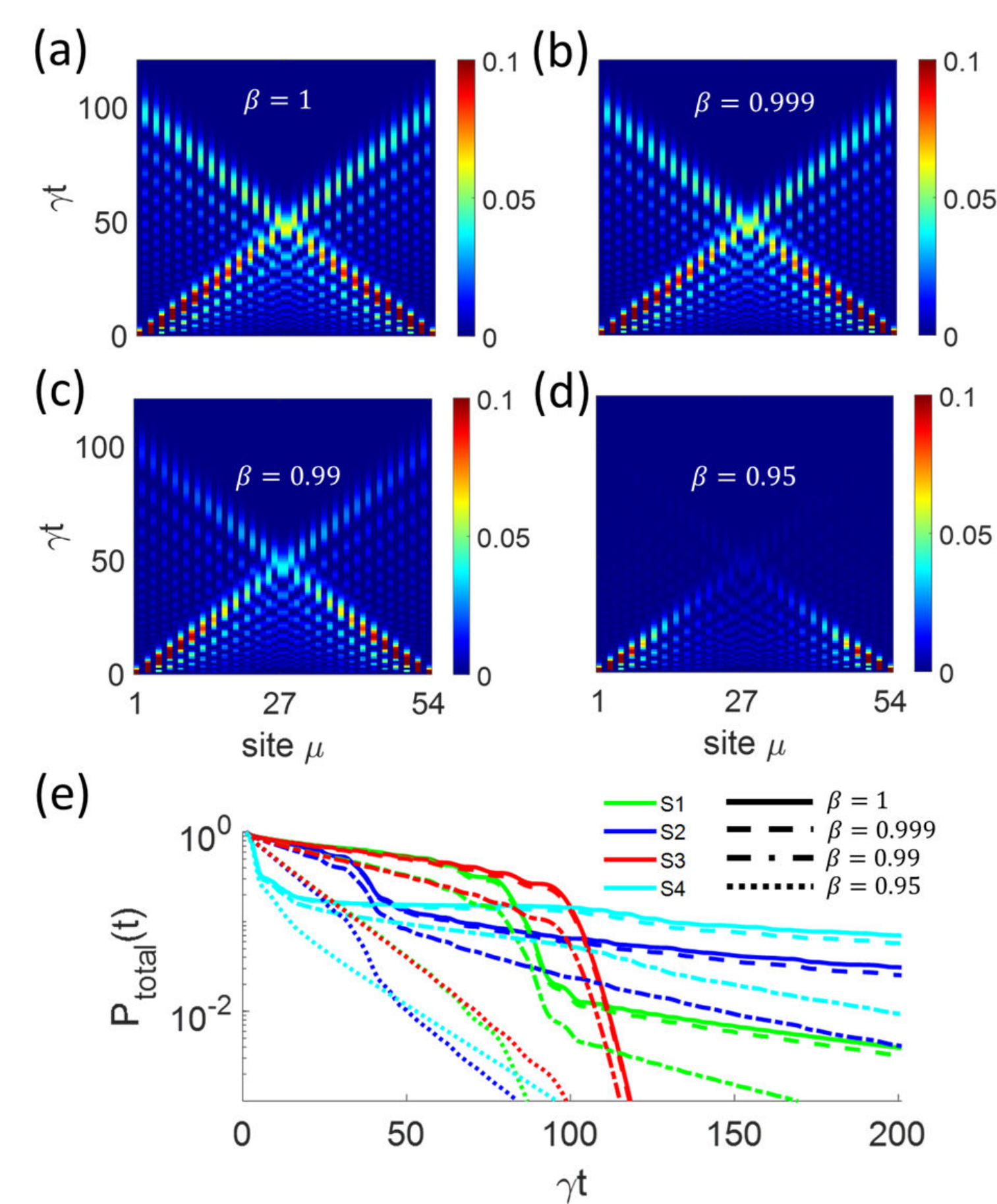}
	\caption{ Experimental considerations in realistic WQED platforms by influence of nonguided mode on excitation transport, characterized by the coupling efficiency $\beta$. (a–d) Excitation dynamics for the S3 configuration at different coupling efficiencies: (a) $\beta=1$, (b) $\beta=0.999$, (c) $\beta=0.99$, and (d) $\beta=0.95$. As $\beta$ decreases, leakage into nonguided modes leads to faster decay of the total population while the underlying propagation pattern remains qualitatively unchanged. (e) Total population dynamics $P_{tot}(t)$ on a logarithmic scale for all structured–directionality patterns: S1 (green), S2 (blue), S3 (red), and S4 (cyan). Solid, dashed, dash-dotted, and dotted lines correspond to $\beta=1$, $0.999$, $0.99$, and $0.95$, respectively. Other parameters, including the interatomic spacing $\xi$ and the number of atoms, are identical to those in Fig. 2.}\label{Fig7}
\end{figure}

Finally we examine the effect of nonideal coupling to the guided modes by introducing a decay channel into the nonguided reservoir. This effect can be quantified through the parameter $\beta = \gamma/(\gamma + \gamma_\text{ng})$ \cite{Lodahl2017}, where $\gamma_{ng}$ denotes dissipation rates from the nonguided modes. Figures \ref{Fig7}(a–d) illustrate the influence of $\beta$ on the excitation transport for the S3 configuration as an example, without loss of generality for other structured patterns. These excitation dynamics behavior shows that as $\beta$ decreases, the overall excitation population weakens, while the spatial transport profile remains qualitatively unchanged. This behavior indicates that the directionality pattern itself is robust against moderate coupling imperfections and continues to govern the excitation flow even when a fraction of emission leaks into the environment.

However, the reduced coupling efficiency leads to faster dissipation of the total excitation before it can propagate across the entire array. For nearly ideal coupling ($\beta = 0.999$ and $0.99$), which correspond respectively to state-of-the-art superconducting-qubit and quantum-dot platforms \cite{Sheremet2023}, the excitation distribution still clearly preserves its structured transport pattern, albeit with lower intensity compared to the perfect case ($\beta = 1$). This confirms that both experimental systems are promising candidates for realizing structured-directionality control in practice.

%========================
%========================

The total population dynamics in Fig. \ref{Fig7}(e) further clarifies this trend. For superconducting-qubit platform at $\beta = 0.999$, all four configurations exhibit decay curves that closely follow the ideal case, showing only minor deviations. For quantum-dot–level coupling at $\beta = 0.99$, the total population decreases slightly faster but still maintains similar qualitative behavior. In contrast, for weaker coupling ($\beta = 0.95$), the total population drops rapidly, demonstrating that strong coupling to the guided modes is essential for observing the collective subradiant features discussed earlier \cite{Tudela2024,Sheremet2023}. These results manifest that the proposed directionality-engineered arrays remain experimentally accessible within current photonic and solid-state platforms.

While $\beta$ directly quantifies radiative loss into nonguided modes, a finite $\eta < 1$ implies that both left- and right-propagating waveguide channels emerge and remain active in the system dynamics. $\beta$ quantifies a coupling efficiency, which generally degrades the overall atomic excitation populations, as shown and studied in Fig. \ref{Fig7}. Meanwhile, $\eta < 1$ allows pattern deformations in various configurations of structures considered in this work, leading to competing propagation pathways. As already discussed in Sec. \ref{sec.spreading}, this competition results in irregular excitation patterns rather than a straightforward suppression of population transport. Nevertheless, the characteristic transport patterns of each structured configuration persist for moderately large chirality. Finite $\eta$ primarily degrades the directionality contrast and renders idealized transport features, such as perfect leap-frog motion, imperfect rather than eliminating the underlying mechanisms.

From an experimental perspective, tuning the chirality factor $\eta$ can be achieved by quantum state-controlled directional spontaneous emissions \cite{Mitsch2014}, where manipulating initialized state populations controls the degrees of chirality in couplings. Even though tuning emitter–waveguide spacing can effectively control the directionality of couplings, for example by positioning atoms within a double-waveguide or geometrically aligning them away from a single waveguide, this would unavoidably lower the coupling efficiency if the spacing is made too far away, limiting the performance of chiral transport and degrading the features presented in this work. We note that tuning the chirality factor $\eta$ and tuning the emitter–waveguide spacing represent different but comparable challenges, as they rely on distinct physical control mechanisms, the state initializations by optical pumping and space manipulations with well control of optical tweezers.

%%%%%%%%%%%%%%
\section{Discussion and Conclusion}\label{sec.discuss}
We demonstrate that introducing site-dependent directionality into waveguide quantum electrodynamics provides a powerful and flexible mechanism to manipulate excitation transport at an atom–nanophotonic interfaces. By systematically engineering the local coupling asymmetry $D_\mu$, we reveal how directionality patterns give rise to a rich variety of dynamical behaviors—ranging from centering and wave-like propagation to leap-frog motion and dispersive dynamics. These regimes emerge purely from the spatial modulation of directionality, without the need to alter the physical waveguide or the interatomic spacing.

Through direct comparison of population dynamics, quasi-steady-state spreads, and spectral properties, we identify the interplay between local asymmetry and collective interference as the essential mechanisms to form long-lived subradiant modes and coherent transport channels. The quasi-steady-state analysis further shows that increasing directionality enhances excitation localization in symmetric configurations, whereas alternating or mixed patterns sustain broader distributions through strong mode mixing. Importantly, the studies including nonguided losses reveal the robustness of strong coupling regime at $\beta \geq 0.99$, showing the feasibility of observing structured-directionality dynamics in existing nanophotonic, quantum-dot, and superconducting-qubit platforms.

Our findings establish structured wQED as a powerful route to manipulate excitation transport and localization, offering new opportunities for programmable quantum state transfer and chiral information routing \cite{Cavazzoni2025, Forero2025} in integrated photonic networks. By treating directionality as a tunable local parameter, this approach effectively maps geometric or multi-waveguide architectures onto a single controllable waveguide platform, thereby bridging diverse experimental implementations within a unified theoretical picture. Beyond the single-excitation regime explored here, extending structured directionality to broader systems such as multiexcitation dynamics \cite{Handayana2025}, non-Markovian couplings \cite{Chang2025}, or nonlinear photonic environments \cite{Ostmann2025} may uncover new forms of correlated photon transport \cite{Mahmoodian2018, Tian2025}, entanglement distributions \cite{Agusti2023}, and decoherence-free subspaces \cite{Karnieli2025}.

%%%%%%%%%%%%%%
\section{Acknowledgments}
 We acknowledge support from the National Science and Technology Council (NSTC), Taiwan, under Grants No. NSTC-112-2119-M-001-007 and No. NSTC-112-2112-M-001-079-MY3, and from Academia Sinica under Grant No. AS-CDA-113-M04.
  
%%%%%%%%%%%%%%%%%%%%%%%%%%%%%%%%%%%

%%%%%%%%%%%%%%%%%%%%%%%%%%%%%%%%%%%%%%%%%%%%%%%%%%%%%%%%

\begin{thebibliography}{99}
%%%%%%%%%%%%%%%%%%%%%%%%%%%%%%%%%%% wQED
\bibitem{Sheremet2023} A. S. Sheremet, M. I. Petrov, I. V. Iorsh, A. V. Poshakinskiy, A. N. Poddubny, Waveguide quantum electrodynamics: collective radiance and photon-photon correlations, Rev. Mod. Phys. {\bf 95}, 015002 (2023).
\bibitem{Vetsch2010} E. Vetsch, D. Reitz, G. Sagu\'e, R. Schmidt, S. T. Dawkins, and A. Rauschenbeutel, Optical Interface Created by Laser-Cooled Atoms Trapped in the Evanescent Field Surrounding an Optical Nanofiber, Phys. Rev. Lett. {\bf 104}, 203603 (2010).
\bibitem{Thompson2013} J. D. Thompson, T. G. Tiecke, N. P. de Leon, J. Feist, A. V. Akimov, M. Gullans, A. S. Zibrov, V. Vuletić, and M. D. Lukin, Coupling a Single Trapped Atom to a Nanoscale Optical Cavity, Science {\bf 340}, 1202 (2013).
\bibitem{Goban2015} A. Goban, C.-L. Hung, J. D. Hood, S.-P. Yu, J. A. Muniz, O. Painter, and H. J. Kimble, Superradiance for Atoms Trapped along a Photonic Crystal Waveguide, Phys. Rev. Lett. {\bf 115}, 063601 (2015).
\bibitem{Corzo2019} N. V. Corzo, J. Raskop, A. Chandra, A. S. Sheremet, B. Gouraud, and J. Laurat, Waveguide-coupled single collective excitation of atomic arrays, Nature {\bf 566}, 359 (2019).
\bibitem{Kim2019} M. E. Kim, T.-H. Chang, B. M. Fields, C.-A. Chen, and C.-L. Hung, Trapping single atoms on a nanophotonic circuit with configurable tweezer lattices, Nat. Commun. {\bf 10}, 1647 (2019).
\bibitem{Dordevic2021} T. Dordevi\ifmmode \acute{c}\else \'{c}\fi{}, P. Samutpraphoot, P. L. Ocola, H. Bernien, B. Grinkemeyer, I. Dimitrova, V. Vuletić, and M. D. Lukin, Entanglement transport and a nanophotonic interface for atoms in optical tweezers, Science {\bf 373}, 1511 (2021). 
\bibitem{Mitsch2014} R. Mitsch, C. Sayrin, B. Albrecht, P. Schneeweiss, and A. Rauschenbeutel, Quantum state-controlled directional spontaneous emission of photons into a nanophotonic waveguide, Nat. Commun. {\bf 5}, 5713 (2014).
\bibitem{Chang2018} D. E. Chang, J. S. Douglas, A. Gonz\'alez-Tudela, C.-L. Hung, H. J. Kimble, \textit{Colloquium}: Quantum matter built from nanoscopic lattices of atoms and photons, Rev. Mod. Phys. {\bf 90}, 031002 (2018).
\bibitem{Kim2021} E. Kim, X. Zhang, V. S. Ferreira, J. Banker, J. K. Iverson, A. Sipahigil, M. Bello, A. Gonz\'alez-Tudela, M. Mirhosseini, and O. Painter, Quantum Electrodynamics in a Topological Waveguide, Phys. Rev. X {\bf 11}, 011015 (2021).
%%%%%%%%%%%%%%%%%%%%%%%%%%%%%%%%%%% guided and photon mediated
\bibitem{Douglas2015} J. S. Douglas, H. Habibian, C.-L. Hung, A. V. Gorshkov, H. J. Kimble, and D. E. Chang, Quantum many-body models with cold atoms coupled to photonic crystals, Nat. Photon. {\bf 9}, 326 (2015).
\bibitem{Solano2017} P. Solano, P. Barberis-Blostein, F. K. Fatemi, L. A. Orozco, and S. L. Rolston, Super-radiance reveals infinite-range dipole interactions through a nanofiber, Nat. commun. {\bf 8}, 1857 (2017).
\bibitem{Tudela2024} A. Gonz\'alez-Tudela, A. Reiserer, J. J. Garc\'ia-Ripoll, and F. J. Garc\'ia-Vidal, Light–matter interactions in quantum nanophotonic devices, Nat. Rev. Phys. {\bf 6}, 166 (2024).
\bibitem{Jen2025} H. H. Jen, Photon-mediated dipole-dipole interactions as a resource for quantum science and technology in cold atoms, Quantum Sci. Technol. {\bf 10}, 023001 (2025).
%%%%%%%%%%%%%%%% subradiant modes
\bibitem{Henriet2019} L. Henriet, J. S. Douglas, D. E. Chang, and A. Albrecht, Critical open-system dynamics in a one-dimensional optical-lattice clock, Phys. Rev. A {\bf 99}, 023802 (2019).
\bibitem{Zhang2019} Y.-X. Zhang and and K. Mølmer, Theory of Subradiant States of a One-Dimensional Two-Level Atom Chain, Phys. Rev. Lett. {\bf 122}, 203605 (2019).
\bibitem{Ke2019} Y. Ke, A. V. Poshakinskiy, C. Lee, Y. S. Kivshar, and A. N. Poddubny, Inelastic Scattering of Photon Pairs in Qubit Arrays with Subradiant States, Phys. Rev. Lett. {\bf 123}, 253601 (2019).
\bibitem{Albrecht2019} A. Albrecht, L. Henriet, A. Asenjo-Garcia, P. B Dieterle, O. Painter, and D. E. Chang, Subradiant states of quantum bits coupled to a one-dimensional waveguide, New J. Phys. {\bf 21}, 025003 (2019).
\bibitem{Needham2019} J. A. Needham, I. Lesanovsky, and B. Olmos, Subradiance-protected excitation transport, New J. Phys. {\bf 21}, 073061 (2019).
\bibitem{Mahmoodian2020} S. Mahmoodian, G. Calajó, D. E. Chang, K. Hammerer, and A. S. Sørensen, Dynamics of Many-Body Photon Bound States in Chiral Waveguide QED, Phys. Rev. X {\bf 10}, 031011 (2020).
\bibitem{Masson2020} S. J. Masson and A. Asenjo-Garcia, Atomic-waveguide quantum electrodynamics, Phys. Rev. Res. {\bf 2}, 043213 (2020). 
\bibitem{Jen2021_bound} H. H. Jen, Bound and subradiant multiatom excitations in an atomic array with nonreciprocal couplings, Phys. Rev. A {\bf 103}, 063711 (2021).
\bibitem{Pennetta2022} R. Pennetta, M. Blaha, A. Johnson, D. Lechner, P. Schneeweiss, J. Volz, and A. Rauschenbeutel, Collective Radiative Dynamics of an Ensemble of Cold Atoms Coupled to an Optical Waveguide, Phys. Rev. Lett. {\bf 128}, 073601 (2022).
\bibitem{Pennetta2022_2} R. Pennetta, D. Lechner, M. Blaha , A. Rauschenbeutel, P. Schneeweiss, and J. Volz, Observation of Coherent Coupling between Super- and Subradiant States of an Ensemble of Cold Atoms Collectively Coupled to a Single Propagating Optical Mode, Phys. Rev. Lett. {\bf 128}, 203601 (2022).
%%%%%%%%%%%%%%%% localization 
\bibitem{Zhong2020} J. Zhong, N. A. Olekhno, Y. Ke, A. V. Poshakinskiy, C. Lee, Y. S. Kivshar, and A. N. Poddubny, Photon-Mediated Localization in Two-Level Qubit Arrays, Phys. Rev. Lett. {\bf 124}, 093604 (2020).
\bibitem{Mirza2017} I. M. Mirza, J. G. Hoskins, and J. C. Schotland, Chirality, band structure, and localization in waveguide quantum electrodynamics, Phys. Rev. A {\bf 96}, 053804 (2017).
\bibitem{Fayard2021} N. Fayard, L. Henriet, A. Asenjo-Garcia, and D. E. Chang, Many-body localization in waveguide quantum electrodynamics, Phys. Rev. Res. {\bf 3}, 033233 (2021).
\bibitem{Wu2024} C.-C. Wu, K.-T Lin, I G. N. Y. Handayana, C.-H. Chien, S. Goswami, G.-D. Lin, Y.-C. Chen, and H. H. Jen, Atomic excitation delocalization at the clean to disorder interface in a chirally-coupled atomic array, Phys. Rev. Research {\bf 6}, 013159 (2024). 
%%%%%%%%%%%%%%%% disorder-induced correlated photon 
\bibitem{Tian2025} G. Tian, L.-L. Zheng, Z.-M. Zhan, F. Nori, and X.-Y. L\"u, Disorder-Induced Strongly Correlated Photons in Waveguide QED, Phys. Rev. Lett. {\bf 135}, 153604 (2025).
%%%%%%%%%%%%%%%% excitation trapping
\bibitem{Handayana2024} I G. N. Y. Handayana, C.-C. Wu, S. Goswami, Y.-C. Chen, and H. H. Jen, Atomic excitation trapping in dissimilar chirally coupled atomic arrays, Phys. Rev. Research {\bf 6}, 013320 (2024).
%%%%%%%%%%%%%%%% correlation
\bibitem{Tudela2013} A. Gonz\'alez-Tudela and D. Porras, Mesoscopic Entanglement Induced by Spontaneous Emission in Solid-State Quantum Optics, Phys. Rev. Lett. {\bf 110}, 080502 (2013).
\bibitem{Mahmoodian2018} S. Mahmoodian, M. \ifmmode \check{C}\else \v{C}\fi{}epulkovskis, S. Das, P. Lodahl, K. Hammerer, and A. S. S\o{}rensen, Strongly Correlated Photon Transport in Waveguide Quantum Electrodynamics with Weakly Coupled Emitters, Phys. Rev. Lett. {\bf 121}, 143601 (2018).
\bibitem{Jeannic2021} H. Le Jeannic, T. Ramos, S. F. Simonsen, T. Pregnolato, Z. Liu, R. Schott, A. D. Wieck, A. Ludwig, N. Rotenberg, J. J. García-Ripoll, and P. Lodahl, Experimental Reconstruction of the Few-Photon Nonlinear Scattering Matrix from a Single Quantum Dot in a Nanophotonic Waveguide, Phys. Rev. Lett. {\bf 126}, 023603 (2021).
\bibitem{Jen2022_correlation} H. H. Jen, Quantum correlations of localized atomic excitations in a disordered atomic chain, Phys. Rev. A {\bf 105}, 023717 (2022).
\bibitem{Handayana2025}  I G. N. Y. Handayana, Y.-L. Tsao, and H. H. Jen, Suppression of Quantum Correlations in a Clean-Disordered Atom-Nanophotonic Interface, Phys. Rev. Research {\bf 7}, 023303 (2025).
%%%%%%%%%%%%%%%% cluster state preparations
\bibitem{Chien2023} C.-H. Chien, S. Goswami, C.-C. Wu, W.-S. Hiew, Y.-C. Chen, and H. H. Jen, Generating scalable graph states in an atom-nanophotonic interface, Quantum Sci. Technol. \textbf{9} 025020 (2024).
\bibitem{Goswami2025} S. Goswami, C.-H. Chien, N. Sinclair, B. Grinkemeyer, S. Bennetts, Y.-C. Chen, and H. H. Jen, Efficient and high-fidelity entanglement in cavity QED without high cooperativity, 	arXiv:2505.02702.
%%%%%%%%%%%%%%%% quantum system
\bibitem{Diehl2010} S. Diehl, A. Tomadin, A. Micheli, R. Fazio, and P. Zoller, Dynamical Phase Transitions and Instabilities in Open Atomic Many-Body Systems, \prl ~{\bf 105}, 015702 (2010).
\bibitem{Kien2005} F. Le Kien, S. Dutta Gupta, K. P. Nayak, and K. Hakuta, Nanofiber-mediated radiative transfer between two distant atoms, \pra~{\bf 72}, 063815 (2005).
%%%%%%%%%%%%%%%% chirality
\bibitem{Pichler2015} H. Pichler, T. Ramos, A. J. Daley, and P. Zoller, Quantum optics of chiral spin networks, Phys. Rev. A {\bf 91}, 042116 (2015).
\bibitem{Lodahl2017} P. Lodahl, S. Mahmoodian, S. Stobbe, A. Rauschenbeutel, P. Schneeweiss, J. Volz, H. Pichler, and P. Zoller, Chiral quantum optics, Nature {\bf 541}, 473 (2017).
\bibitem{Gardiner1993} C. W. Gardiner, Driving a quantum system with the output field from another driven quantum system, Phys. Rev. Lett. {\bf 70}, 2269 (1993).
\bibitem{Carmichael1993} H. J. Carmichael, Quantum trajectory theory for cascaded open systems, Phys. Rev. Lett. {\bf 70}, 2273 (1993).
\bibitem{Stannigel2012} K. Stannigel, P. Rabl, and P. Zoller, Driven-dissipative preparation of entangled states in cascaded quantum-optical networks, New J. Phys. {\bf 14}, 063014 (2012).
\bibitem{Jen2020_collective} H. H. Jen, Collective single excitation dynamics in a chirally coupled atomic chain, J. Phys. B: At. Mol. Opt. Phys. {\bf 53}, 205501 (2020).
\bibitem{Holzinger2024} R. Holzinger, J. S. Peter, S. Ostermann, H. Ritsch, and S. Yelin, Harnessing quantum emitter rings for efficient energy transport and trapping, Optica Quantum {\bf 2}, 57 (2024).
\bibitem{Diepen2025} C. J. van Diepen, V. Angelopoulou, O. A. D. Sandberg, A. Tiranov, Y. Wang, S. Scholz, A. Ludwig, A. S. S\o{}rensen, and P. Lodahl, Resonant energy transfer and collectively driven emitters in waveguide QED, Phys. Rev. Res. {\bf 7}, 033169 (2025).
\bibitem{Chen2025} W. Chen, G.-D. Lin and H.H. Jen, Excitation transfer and many-body dark states in waveguide quantum electrodynamics, Phys. Rev. A {\bf 112}, 043713 (2025).
\bibitem{Yu2025} Y.-T. Yu, I G. N. Y. Handayana, W. Chen, and H. H. Jen, Fast and High Excitation Transport in Waveguide Quantum Electrodynamics, arXiv:2506.11885.
\bibitem{Forero2025} D. G. Su\'arez-Forero, M. J. Mehrabad, C. Vega, A. Gonz\'alez-Tudela, and M. Hafezi, Chiral Quantum Optics: Recent Developments and Future Directions, Phys. Rev. X {\bf 6}, 020101 (2025).
%%%%%%%%%%%%%%%% structure wQED
\bibitem{Yang2022} M. Yang, L. Wang, X. Wu, H. Xiao, D. Yu, L. Yuan, and X. Chen, Concentrated subradiant modes in a one-dimensional atomic array coupled with chiral waveguides, Phys. Rev. A {\bf 106}, 043717 (2022).
\bibitem{Lee2025} C.-Y. Lee, K.-T. Lin, G.-D. Lin, and H. H. Jen, Controlling Excitation Localization in Waveguide QED Systems, to be published in Phys. Rev. Research (2025).
%%%%%%%%%%%%%%%%%%%%%%%%%%%%%%%%%% SQ and QD and atoms
\bibitem{Roushan2017} P. Roushan, C. Neill, A. Megrant, Y. Chen, R. Babbush, R. Barends, B. Campbell, Z. Chen, B. Chiaro, A. Dunsworth, {\it et. al.}, Chiral ground-state currents of interacting photons in a synthetic magnetic field, Nat. Phys. {\bf 13}, 146 (2017).  
\bibitem{Wang2019} D.-W. Wang, C. Song, W. Feng, H. Cai, D. Xu, H. Deng, H. Li, D. Zheng, X. Zhu, H. Wang, {\it et. al.}, Synthesis of antisymmetric spin exchange interaction and chiral spin clusters in superconducting circuits, Nat. Phys. {\bf 15}, 382 (2019).  
\bibitem{Luxmoore2013} I. J. Luxmoore, N. A. Wasley, A. J. Ramsay, A. C. T. Thijssen, R. Oulton, M. Hugues, S. Kasture, V. G. Achanta, A. M. Fox, and M. S. Skolnick, Interfacing Spins in an InGaAs Quantum Dot to a Semiconductor Waveguide Circuit Using Emitted Photons, Phys. Rev. Lett. {\bf 110}, 037402 (2013). 
\bibitem{Arcari2014} M. Arcari, I. S\"{o}llner, A. Javadi, S. Lindskov Hansen, S. Mahmoodian, J. Liu, H. Thyrrestrup, E. H. Lee, J. D. Song, S. Stobbe, and P. Lodahl, Near-Unity Coupling Efficiency of a Quantum Emitter to a Photonic Crystal Waveguide, Phys. Rev. Lett. {\bf 113}, 093603 (2014).
\bibitem{Yalla2014} R. Yalla, M. Sadgrove, K. P. Nayak, and K. Hakuta, Cavity Quantum Electrodynamics on a Nanofiber Using a Composite Photonic Crystal Cavity, Phys. Rev. Lett. {\bf 113}, 143601 (2014).
\bibitem{Sollner2015} I. Söllner, S. Mahmoodian, S. L. Hansen, L. Midolo, A. Javadi, G. Kiršanskė, T. Pregnolato, H. El-Ella, E. H. Lee, J. D. Song, {\it et. al.}, Deterministic photon–emitter coupling in chiral photonic circuits, Nat. Nanotechnol. {\bf 10}, 775 (2015).  
\bibitem{Morrissey2009} M. J. Morrissey, K. Deasy, Y. Wu, S. Chakrabarti, and S. Nic Chormaic, Tapered optical fibers as tools for probing magneto-optical trap characteristics, Rev. Sci. Instrum. {\bf 80}, 053102 (2009).
%%%%%%%%%%%%%%%%%%%%%%%%%%%%%%%%%% quantum network 
\bibitem{Sunami2025} S. Sunami, S. Tamiya, R. Inoue, H. Yamasaki, and A. Goban, Scalable Networking of Neutral-Atom Qubits: Nanofiber-Based Approach for Multiprocessor Fault-Tolerant Quantum Computers, PRX Quantum {\bf 6}, 010101 (2025).
%%%%%%%%%%%%%%%%%%%%%%%%%%%%%%%%%% near-field effect 
\bibitem{Kuraptsev2020} A. S. Kuraptsev and I. M. Sokolov, Incomplete spontaneous decay in a waveguide caused by polarization selection, Phys. Rev. A {\bf 101}, 053852 (2020).
%%%%%%%%%%%%%%%%%%%%%%%%%%%%%%%%%% Effective hamiltonan 
\bibitem{Lehmberg1970} R. H. Lehmberg, Radiation from an $N$-Atom System. I. General Formalism, Phys. Rev. A {\bf 2}, 883 (1970).
%%%%%%%%%%%%%%%%%%%%%%%%%%%%%%%%%% Non hermitian 
\bibitem{Ashida2020} Y. Ashida, Z. Gong, and M. Ueda, Non-Hermitian physics, Advances in Physics {\bf 69}, 249 (2020).
%%%%%%%%%%%%%%%%%%%%%%%%%%%%%%%%%% Dimerization 
\bibitem{Poshakinskiy2021} A. V. Poshakinskiy and A. N. Poddubny, Dimerization of Many-Body Subradiant States in Waveguide Quantum Electrodynamics, Phys. Rev. Lett. {\bf 127}, 173601 (2021).
%%%%%%%%%%%%%%%%%%%%%%%%%%%%%%%%%% Routing 
\bibitem{Cavazzoni2025} S. Cavazzoni, G. Ragazzi, P. Bordone, and M. G. A. Paris, Perfect chiral quantum routing, Phys. Rev. A {\bf 111}, 032439 (2025).
%%%%%%%%%%%%%%%%%%%%%%%%%%%%%%%%%% Future work
\bibitem{Chang2025} Y. Chang, Non-Markovian multiphoton chiral dynamics with giant systems, Commun. Phys. {\bf 8}, 385 (2025).
\bibitem{Ostmann2025} M. Ostmann, J. Nunn, and A. E. Jones, Nonlinear photonic architecture for fault-tolerant quantum computing, arXiv:2510.06890.
\bibitem{Agusti2023} J. Agust\'i, X. H. H. Zhang, Y. Minoguchi, and P. Rabl, Autonomous Distribution of Programmable Multiqubit Entanglement in a Dual-Rail Quantum Network, Phys. Rev. Lett. {\bf 131}, 250801 (2023).
\bibitem{Karnieli2025} A. Karnieli, O. Tziperman, C. Roques-Carmes, and S. Fan, Decoherence-free many-body Hamiltonians in nonlinear waveguide quantum electrodynamics, Phys. Rev. Res. {\bf 7}, L012014 (2025).

\end{thebibliography}
\end{document}